# Signature of chiral anomaly in a graphene-based heterostructure

Ivan S. Sokolov,[a] Dmitry V. Averyanov,[a] Oleg E. Parfenov,[a] Sergey P. Lebedev,[b] Alexander A. Lebedev,[b] Olesya O. Kapitanova,[c] Alexey N. Mihalyuk,[d,e] Oleg A. Kondratev,[a] Andrey M. Tokmachev,[a] Vyacheslav G. Storchak[*,a]

[a] National Research Center "Kurchatov Institute", Kurchatov Sq. 1, 123182 Moscow, Russia

[b] Ioffe Institute, 194021 St. Petersburg, Russia

[c] Lomonosov Moscow State University, 119991 Moscow, Russia

[d] Institute of High Technologies and Advanced Materials, Far Eastern Federal University, 690950 Vladivostok, Russia

[e] Institute of Automation and Control Processes FEB RAS, 690041 Vladivostok, Russia

**Abstract**

Chiral anomaly is a fundamental manifestation of topological matter caused by chirality imbalance. It can be of great practical value by providing a mechanism for dissipationless transport. The phenomenon is highly sensitive to the system dimensionality, disappearing in the 2D limit. Different schemes have been proposed to observe the chiral anomaly in graphene-based systems but experimental evidence has been lacking. Here, we synthesized two types of the Gr/Sr/SiC nanostructure – with 2D and 3D, bulk-like, layers of Sr – the difference being the amount of Sr intercalated under graphene. In contrast to the 2D material, 3D Gr/Sr/SiC exhibits a signature of the chiral anomaly – quadratic negative magnetoresistance in parallel magnetic and electric fields. Also, the functional form of the temperature dependence of the magnetoresistance agrees with the chiral anomaly. In addition, the 3D heterostructure exhibits suppressed conductivity and quantum oscillations in graphene. These effects may be associated with the graphene layer incorporated into a complex 3D heterostructure. Hopefully, the discovered signature of chiral anomaly in the built-in graphene would spur research on the phenomenon in topological nanomaterials.

[*] Corresponding author. *E-mail address:* vgstorchak9@gmail.com (Vyacheslav Storchak)

## 1. Introduction

The discoveries of topological phases have opened vast opportunities in materials science [1-3], bringing forward a wide range of applications, from spintronic devices to chemical catalysis. A key property of fermions in topological phases is their chirality [4-6]. The chirality imbalance, which can be induced by electromagnetic fields, gives rise to such a fundamental characteristic of topological matter as the chiral anomaly (CA) [7]. Its most prominent signature is the chiral magnetic effect – the emergence of negative magnetoresistance (MR) in parallel magnetic and electric fields [8-10]. This is not the only manifestation of the CA – its effects can be detected by THz spectroscopy [11] and thermal conductivity measurements [12]. The fundamental importance of the CA is in being a sensitive probe of the topological electronic structure. Its technological uses are probably even of higher significance: the dissipationless character of the topological chiral current paves the way to low-power electronics. The CA concept inspires a variety of applications, including qubit realization for quantum computing [13], lenses for subwavelength imaging [14], or THz detectors [15]. The utility of the CA has been demonstrated experimentally by design of highly efficient topological transistor devices based on the Dirac semimetal $PtSe_2$ [16] and the Weyl semiconductor Te [17,18]. The current state of affairs is that the CA observation and use in devices rely upon bulk-like materials such as rather thick flakes whereas modern electronics requires nanomaterials. An outstanding question is whether the CA studies and applications can be extended to the 2D limit.

The fundamental problem with the 2D limit is that CA is associated with odd spatial dimensions [19,20]. Accordingly, the CA signatures should disappear as the film of the topological material becomes ultrathin. For example, 4-nm-thick $PtSe_2$ [16] and 12-nm-thick Te [17] lack the negative MR observed in thicker flakes and associated with CA. A layer-dependent study of silicene-based $SrSi_2$ demonstrates that in ultrathin films, CA gives way to weak localization effects [21]. Recently, an experimental realization of chiral Landau levels (giving rise to CA) has been demonstrated in a 2D photonic honeycomb system [22]. The potential emergence of CA is analyzed theoretically for various quasi-2D systems, from honeycomb-lattice based ballistic conductors [23] to 2D phononic crystal slabs [24]. Most attention has been paid to graphene-related systems [25-30]. In fact, CA in graphene had been discussed well before the first experiments on graphene [20]. In particular, CA is theorized to emerge because of finite-size effects, in graphene ribbons [27-29] or a graphene plate [30]. The general idea is that to induce CA certain symmetries should be broken. Thus, pristine or slightly modified graphene are unlikely to demonstrate CA; instead, one should devise a strategy for radical alteration of graphene.

The integrity of graphene can be broken chemically. One way is to introduce defects: they are known to tune the CA in other topological materials, leading to an increase in the negative longitudinal MR [31]. Defects and chemical modifications of graphene can drastically change its transport properties and induce a negative MR. However, this negative MR emerges in out-of-plane magnetic field [32-36]; the anisotropy is such that the MR in in-plane magnetic field is positive [35,36] and thus can hardly be associated with the CA. An alternative approach would be to deform graphene using a highly uneven surface and trigger finite-size effects. In particular, metal intercalation into epitaxial graphene/SiC is regarded as a way to modulate graphene [37]. In the context of coupling with graphene, rare earths are probably among the most studied metals, Eu [38-45] and Gd [46-50] in particular. Eu intercalation may produce either a regular flat structure [44] or nano-clusters of Eu located on the vertices of a Moiré pattern [42,43]. The interest to the Eu- and Gd-based materials stems from their advantageous magnetic properties. However, the same magnetic properties are a hindrance in a study of the CA. Alternatively, one can employ the well-established correspondence between magnetic Eu and nonmagnetic Sr compounds [51,52]: as a rule, Eu and Sr compounds are isostructural, with close lattice parameters. In particular, it applies to the graphite intercalation compounds $EuC_6$ and $SrC_6$ [53] and their monolayers [44].

However, such an approach may be insufficient to induce CA because it may not lift the dimensionality restrictions. Stronger modifications are needed. A 2D system can be embedded into a complex 3D structure to exhibit CA. Such a strategy may fly. This idea is supported by the observation of a CA signature in a 2D electron gas at the $LaVO_3/KTaO_3$ interface with a rather thick conducting channel [54]. Similarly, graphene can be incorporated into a 3D heterostructure.

Here, we produce and compare two types of materials formed by Sr intercalation under epitaxial graphene on SiC. One is a 2D structure – a monolayer of Sr between graphene and SiC. The other couples graphene to a layer of bulk Sr. The amount of intercalated Sr makes a strong impact on the electron transport. The 3D material is discovered to exhibit a signature of the CA – a negative longitudinal MR with characteristic magnetic field and temperature dependences.

## 2. Experimental section

### *2.1. Synthesis*

The synthesis comprised production of epitaxial graphene on SiC and its intercalation with Sr. The first step was carried out using the SiC surface sublimation method [55]. The employed wafers of 4H-SiC(0001) (TanKeBlue Semiconductor Co., Ltd.) were semi-insulating, with a miscut angle below 0.25°. The surface of the SiC substrate was polished both chemically and mechanically. The layer of epitaxial graphene was produced in a high-frequency induction heating system employing a graphite crucible: the substrate was gradually heated under Ar atmosphere (750 ± 20 Torr) up to 1730 ± 20 °C (100 °C/min) and then annealed for 5 min. The SiC wafer was then transferred into a Riber Compact system for molecular beam epitaxy (MBE); its growth chamber maintained the base pressure below $10^{-10}$ Torr. The SiC surface was annealed for 10 min at 400 °C and then cooled down to 150 °C. Gr/Sr/SiC was synthesized by deposition of Sr supplied from a Knudsen cell effusion source heated up to 240 °C. According to a Bayard-Alpert ionization gauge, the partial pressure of Sr was kept at $1 \cdot 10^{-8}$ Torr. Two characteristic samples of Gr/Sr/SiC were produced. One contains a monolayer of Sr corresponding to the $(\sqrt{3} \times \sqrt{3})R30°$ phase; it is denoted as 2D-Gr/Sr/SiC. The other accommodates a film of bulk Sr phase (the time of Sr deposition is increased by 4 times with respect to 2D-Gr/Sr/SiC); it is denoted as 3D-Gr/Sr/SiC. To prevent oxidation of the samples, they were capped at room temperature with a 20-nm layer of amorphous $SiO_x$.

### *2.2. Characterization*

The atomic structure of the samples was characterized *in situ* employing reflection high-energy electron diffraction (RHEED). The diffractometer was equipped with a kSA 400 analytical system. *Ex situ* analysis of the structure was carried out using grazing incidence X-ray diffraction (GIXRD). GIXRD scans were produced in a Rigaku SmartLab 9 kW diffractometer with a Cu $K_{\alpha 1}$ X-ray source ($\lambda$ = 1.54056 Å). Micro Raman spectra were collected with a confocal scanning Raman microscope Horiba LabRAM HR Evolution equipped with a linearly polarized laser (532 nm, 50 mW laser power, a spot size of ~0.72 μm). Unpolarized detection was used to provide a significant signal-to-noise ratio. The lateral electron transport in Gr/Sr/SiC was studied using a LakeShore 9709A system. Four-contact measurements employed square samples with a lateral size of 5 mm; they adhered to the ASTM standard F76. The electrical contacts were made by deposition of Ag-Sn-Ga eutectic alloy; their ohmicity was established by analysis of I-V characteristic curves.

### *2.3. Computational details*

The PBE calculations were carried out with the VASP package employing the projector-augmented wave approach. The substrate was simulated by a slab consisting of six bilayers of 4H-SiC(0001) with a PBE-optimized bulk lattice constant ($a$ = 3.068 Å). The positions of the atoms in the two bottom SiC bilayers were fixed and the rest was relaxed until the atomic forces were below 0.01 eV/Å. The dangling bonds were passivated by H atoms. The vacuum space between the slabs was set to 1.5 nm. The Gr/Sr/SiC systems were simulated using a $3 \times 3$ SiC supercell. The plane-wave cut-off energy was set to 400 eV; the Brillouin zone integration was carried out with a Gaussian smearing of 10 meV. A $6 \times 6 \times 1$ Γ-centered $k$-point mesh was used to sample the Brillouin zone. The scalar relativistic effect and the spin-orbit coupling were taken into account.

## 3. Results and discussion

### *3.1. Synthesis and structural characterization*

The surface of SiC is well known to produce a layer of epitaxial graphene [56,57]. The graphene layer on SiC has certain advantages over exfoliated graphene, stemming from its rather uniform synthesis, wafer-scale monocrystallinity, and compatibility with semiconductor technologies. Epitaxial graphene is suggested as a potential material in electronic and spintronic devices [44,58]. In the present study, we used wafers of 4H-SiC; the layer of epi-graphene was produced by the surface sublimation method. The resulting Gr/SiC material is a natural system for metal intercalation [37,57]. However, one should take into account a rather complex structure of the surface layer in Gr/SiC; in particular, there is a graphene-like buffer layer below the layer of epi-graphene. Although the buffer layer is covalently bound to the SiC bulk, metal atoms can still be intercalated under this layer. Accordingly, during synthesis of Gr/Sr/SiC, the substrate was kept at a low temperature of 150 °C, to prevent Sr intercalation under the buffer layer. Further decrease of temperature would harm the main process of Sr intercalation under epi-graphene. Intercalation of metal atoms can be carried out using different techniques. Our choice here is MBE, a versatile technique that has been useful in controlled synthesis of graphene/metal systems [41,44,47].

The substrate temperature is not the only parameter in the intercalation process. Another important parameter is the amount of deposited Sr. It controls whether a 2D or 3D phase of Sr is formed under graphene. For our study we have chosen two characteristic systems. The first, 2D-Gr/Sr/SiC, corresponds to a monolayer of the graphite intercalation compound $SrC_6$. According to DFT calculations, such a structure has the lowest energy among different positions of Sr atoms in the Gr/SiC system [44]. CA is not expected in this 2D material; instead, it can serve as a reference sample. The other system, 3D-Gr/Sr/SiC, contains a large volume of Sr. If judged by the amount of deposited Sr, 3D-Gr/Sr/SiC has 4 times more Sr than 2D-Gr/Sr/SiC. The additional Sr does not form any chemical compound with graphene (or SiC) and, thus, is expected to crystallize as elemental Sr, similar to Eu intercalated into Gr/SiC [42,43]. The idea is that this Sr crystal would bring the third dimension into the system to induce CA. The amount of Sr that can be used for this purpose is limited: further intercalation results in deterioration of the material crystallinity. The layer of graphene protects a metal from oxidative environment [59]. However, the high chemical reactivity of Sr may still result in samples degradation due to air oxygen and possibly water vapors. Therefore, to be sure that the samples are not oxidized, they are covered with $SiO_x$, an amorphous nonmagnetic insulator that does not affect the results of *ex situ* experiments.

The atomic structure of both the Gr/Sr/SiC materials was characterized in the MBE growth chamber using RHEED. This technique is sensitive to the surface layer. Figure 1a demonstrates a RHEED scan for the 2D-Gr/Sr/SiC phase. The emerging bright reflexes correspond to a regular structure of Sr atoms, namely, a $(\sqrt{3} \times \sqrt{3})R30°$ superstructure. This sort of graphene/metal phase is not uncommon; in particular, it is observed for Eu adsorption and intercalation [38,44]. It can be described as a system where each third hexagon of the graphene lattice accommodates a Sr atom. In the process of intercalation, an intermediate $(2 \times 2)$ phase may appear, as established for Eu [39]; however, we did not observe such a phase for Sr. The addition of the third dimension to the Sr layer changes the structure and, naturally, the RHEED image. The most important distinction is that the RHEED scan for 3D-Gr/Sr/SiC acquires a streaky pattern (Figure 1b). It indicates the formation of bulk Sr layer. Dimmed reflexes of the SiC substrate provide another indication of the significant increase in the thickness of the Sr layer. The reflexes of Sr in Figures 1a and b appear at the same positions: it is expected because the $(\sqrt{3} \times \sqrt{3})R30°$ superstructure in 2D-Gr/Sr/SiC serves as a seeding layer setting the crystal orientation (111) for bulk Sr in 3D-Gr/Sr/SiC. The distance between the neighboring Sr atoms is 4.26 Å in the $(\sqrt{3} \times \sqrt{3})R30°$ superstructure and 4.30 Å in the (111) planes of bulk Sr.

Further structural characterization is provided by GIXRD studies. Figures 1c and d demonstrate signals from all the constituents of the heterostructures, Gr, Sr and SiC. The signals from Sr in 2D-Gr/Sr/SiC are identified as coming from the $(\sqrt{3} \times \sqrt{3})R30°$ superstructure detected in the RHEED scan. A similar GIXRD spectrum is reported for Gr/Eu/SiC [44]. The significant increase in the Sr peaks intensity in Figure 1d confirms the evolution from the 2D Sr superstructure to a 3D Sr layer. In both materials, no side phases are detected. The quality of the graphene layer is probed using Raman spectroscopy. The spectra include significant contributions from the SiC substrate. To extract the signals from the graphene layer, the separately measured spectrum of the substrate is subtracted, as in Ref. [60]. Figures 1e and f demonstrate that the Sr intercalation does not affect significantly the graphene layer. In particular, the increase of the D peak is moderate suggesting that the amount of defects is rather low.

*3.2. Electron transport properties*

The atomic structure characterization of the Gr/Sr/SiC materials does not provide a comprehensive picture of changes occurring in graphene due to Sr intercalation. In this regard, electron transport measurements may be particularly useful: the conductivity of the heterostructure is strongly dominated by the graphene layer (SiC is an insulator), especially at low temperature, making the lateral electron transport sensitive to the state of graphene. Figure 2 shows temperature dependence of sheet resistance in the two materials, 2D- and 3D-Gr/Sr/SiC. The general form of the $R_S(T)$ curves is similar – resistivity increase with temperature. However, there are important differences. The resistance of 3D-Gr/Sr/SiC is noticeably higher than that of 2D-Gr/Sr/SiC, despite addition of Sr atoms that are capable of graphene doping (according to our Hall effect measurements, the carrier concentration in 3D-Gr/Sr/SiC is about 5 times higher than that in 2D-Gr/Sr/SiC). The reduced conductivity of 3D-Gr/Sr/SiC may be associated with graphene incorporation into the 3D structure and/or additional defects witnessed by Raman spectroscopy (Figure 1e and f). The same conclusion can be drawn from the lower residual-resistance ratio (RRR) in 3D-Gr/Sr/SiC as compared to 2D-Gr/Sr/SiC.

Another way to probe the graphene layer is to analyze quantum (Shubnikov-de Haas) oscillations that are commonly observed in graphene. Figures 3a and b demonstrate the oscillatory contributions to the sheet resistance in 2D-Gr/Sr/SiC and 3D-Gr/Sr/SiC, respectively. The comparison shows that the amplitude of oscillations in 3D-Gr/Sr/SiC is an order of magnitude lower. Moreover, the oscillations in 3D-Gr/Sr/SiC comprise only two periods. Analysis of the oscillations suggests that the quantum carrier mobilities in the two Gr/Sr/SiC systems differ significantly, from 850 $cm^2 V^{-1} s^{-1}$ in 2D-Gr/Sr/SiC to 120 $cm^2 V^{-1} s^{-1}$ in 3D-Gr/Sr/SiC. The difference between the materials is even more pronounced in the Hall resistance measurements: $R_{xy}$ in 2D-Gr/Sr/SiC (Figure 3c) exhibits well-developed quantum oscillations but no such oscillations are detected in the magnetic field dependence of $R_{xy}$ in 3D-Gr/Sr/SiC (Figure 3d). All these data point at significant changes in the electronic structure of graphene due to its coupling to 3D Sr.

The presence of CA is usually detected in measurements of MR, by emergence of a negative longitudinal MR with characteristic temperature and field dependences. Figure 4a demonstrates dependence of low-temperature MR in 2D-Gr/Sr/SiC on magnetic field parallel to the current. On the one hand, the MR is negative, as in the case of CA. On the other hand, its magnetic field dependence is linear, in contrast to theoretical predictions of a quadratic magnetic field dependence of the chiral magnetic effect [10,61,62]. Moreover, the MR anisotropy for in-plane magnetic fields is rather low, the MR in magnetic fields perpendicular to the current being even more negative. Therefore, the negative longitudinal MR in 2D-Gr/Sr/SiC is not associated with a CA. The negative longitudinal MR in 3D-Gr/Sr/SiC is different (Figure 4b); it comprises two contributions. The first (negative) contribution has a quadratic form, in accord with the expected magnetic field dependence of the MR due to the chiral magnetic effect. The other, relatively small contribution present at low magnetic fields is associated with weak antilocalization. A similar fusion of CA and weak antilocalization has been observed in other topological materials, such as $SrSi_2$ [21] and $Cd_3As_2$ [63,64]. Regretfully, it is difficult to compare the negative longitudinal MR in 3D-Gr/Sr/SiC with that in a magnetic field perpendicular to the current because the latter is strongly dominated by localization effects.

The difference between the materials can be probed by model DFT calculations (Figure 5). They show that the Sr intercalation to form 2D-Gr/Sr/SiC destroys the Dirac dispersion of Gr/SiC: a hybridization gap opens up at the Fermi level to form a rather complex band structure. In contrast, the Dirac dispersion is restored in 3D-Gr/Sr/SiC.

Figure 4b suggests that 3D-Gr/Sr/SiC may support CA. However, to back up this hypothesis a more quantitative analysis is necessary. First, we fitted the magnetoconductance at different temperatures by quadratic functions $\sigma = \sigma_0 + aB^2$. An example of such a fit, for $\sigma$ at 1.8 K, is provided in Figure 6a. As a result, we extracted the coefficients $a$ for a set of 9 temperatures. The coefficients drop by an order of magnitude with temperature but do not approach zero; it may point at some residual contribution to the MR. The chiral magnetic effect is expected to decrease with temperature; more precisely, it is expected to be proportional to $(T^2 + \mu^2/\pi^2)^{-1}$, where $\mu$ is the chiral chemical potential. We fitted the variation in the normalized coefficient $a$ using this function (Figure 6b). The optimized value of $\mu$ is rather low, about 2.8 meV (*cf.* 9 meV in $ZrTe_5$, an archetypal topological material). In general, the fit is quite accurate. It means that both magnetic field and temperature dependences of the longitudinal MR are in agreement with the presence of a CA.

## 4. Conclusion

Weyl and Dirac semimetals can be considered as 3D analogs of graphene [65]. One of the most prominent manifestations of their topological nature is CA. Nowadays, CA is not only an object of fundamental research but it also finds practical applications due to particularly advantageous properties of the chiral currents. Therefore, the prospect of extending the CA physics from topological semimetals to their close relative graphene seems very enticing. A number of theoretical studies are devoted to graphene-based systems hosting a CA; yet, experimental detection of such a CA remained elusive. The problem is general for 2D systems. We designed a 3D heterostructure Gr/Sr/SiC containing a graphene layer and set it against its 2D counterpart. Electron transport measurements reveal an unconventional longitudinal MR arising in 3D-Gr/Sr/SiC, different from that in 2D-Gr/Sr/SiC. The MR in the 3D structure is negative and its magnetic field dependence is quadratic, as expected in the case of a CA. Moreover, temperature dependence of the effect is also consistent with a CA. It should be stressed that these results cannot be considered as a proof of CA in 3D-Gr/Sr/SiC. Indeed, MR data would not be enough to make definitive statements on the presence of CA even in the case of 3D topological semimetals; less so in such a complex nanomaterial where the mechanism of potential chiral imbalance is unclear. Therefore, we can only state that we observe a signature of the CA, the same as many other works on the CA in 3D materials [7,66-68]. However, we hope that the present study would bring attention to the prospects of CA generation in topological nanomaterials.

## Acknowledgments

This work was supported by NRC “Kurchatov Institute” (structural characterization) and the Russian Science Foundation [grants No. 25-79-10047 (synthesis) and No. 24-19-00038 (studies of electron transport)]. The measurements were carried out using equipment of the resource centers of electrophysical and laboratory X-ray techniques at NRC “Kurchatov Institute”. The Raman spectroscopy measurements were carried out using the equipment of the Moscow State University (the Equipment renovation program, National project “Science”).

## References

[1] P. Narang, C.A.C. Garcia, C. Felser, The topology of electronic band structures, Nat. Mater. 20 (2021) 293.

[2] B.J. Wieder, B. Bradlyn, J. Cano, Z. Wang, M.G. Vergniory, L. Elcoro, A.A. Soluyanov, C. Felser, T. Neupert, N. Regnault, B.A. Bernevig, Topological materials discovery from crystal symmetry, Nat. Rev. Mater. 7 (2022) 196.

[3] B.A. Bernevig, C. Felser, H. Beidenkopf, Progress and prospects in magnetic topological materials, Nature 603 (2022) 41.

[4] M.Z. Hasan, G. Chang, I. Belopolski, G. Bian, S.-Y. Xu, J.-X. Yin, Weyl, Dirac, and high-fold chiral fermions in topological quantum matter, Nat. Rev. Mater. 6 (2021) 784.

[5] Q. Ma, S.-Y. Xu, C.-K. Chan, C.-L. Zhang, G. Chang, Y. Lin, W. Xie, T. Palacios, H. Lin, S. Jia, P.A. Lee, P. Jarillo-Herrero, N. Gedik, Direct optical detection of Weyl fermion chirality in a topological semimetal, Nat. Phys. 13 (2017) 842.

[6] N.B.M. Schröter, S. Stolz, K. Manna, F. de Juan, M.G. Vergniory, J.A. Krieger, D. Pei, T. Schmitt, P. Dudin, T.K. Kim, C. Cacho, B. Bradlyn, H. Borrmann, M. Schmidt, R. Widmer, V.N. Strocov, C. Felser, Observation and control of maximal Chern numbers in a chiral topological semimetal, Science 369 (2020) 179.

[7] N.P. Ong, S. Liang, Experimental signatures of the chiral anomaly in Dirac-Weyl semimetals, Nat. Rev. Phys. 3 (2021) 394.

[8] J. Xiong, S.K. Kushwaha, T. Liang, J.W. Krizan, M. Hirschberger, W. Wang, R.J. Cava, N.P. Ong, Evidence for the chiral anomaly in the Dirac semimetal $Na_3Bi$, Science 350 (2015) 413.

[9] X. Huang, L. Zhao, Y. Long, P. Wang, D. Chen, Z. Yang, H. Liang, M. Xue, H. Weng, Z. Fang, X. Dai, G. Chen, Observation of the chiral-anomaly-induced negative magnetoresistance in 3D Weyl semimetal TaAs, Phys. Rev. X 5 (2015) 031023.

[10] Q. Li, D.E. Kharzeev, C. Zhang, Y. Huang, I. Pletikosić, A.V. Fedorov, R.D. Zhong, J.A. Schneeloch, G.D. Gu, T. Valla, Chiral magnetic effect in $ZrTe_5$, Nat. Phys. 12 (2016) 550.

[11] B. Cheng, T. Schumann, S. Stemmer, N.P. Armitage, Probing charge pumping and relaxation of the chiral anomaly in a Dirac semimetal, Sci. Adv. 7 (2021) eabg0914.

[12] D. Vu, W. Zhang, C. Şahin, M.E. Flatté, N. Trivedi, J.P. Heremans, Thermal chiral anomaly in the magnetic-field-induced ideal Weyl phase of $Bi_{1-x}Sb_x$, Nat. Mater. 20 (2021) 1525.

[13] D.E. Kharzeev, Q. Li, The chiral qubit: quantum computing with chiral anomaly, arXiv:1903.07133 (2019).

[14] S. Tchoumakov, J. Cayssol, A.G. Grushin, Three-dimensional chiral Veselago lensing, Phys. Rev. B 105 (2022) 075309.

[15] S. Kaushik, D.E. Kharzeev, E.J. Philip, Chiral magnetic photocurrent in Dirac and Weyl materials, Phys. Rev. B 99 (2019) 075150.

[16] J. Chen, T. Zhang, J. Wang, N. Zhang, W. Ji, S. Zhou, Y. Chai, Field-effect chiral anomaly devices with Dirac semimetal, Adv. Funct. Mater. 31 (2021) 2104192.
[17] J. Chen, T. Zhang, J. Wang, L. Xu, Z. Lin, J. Liu, C. Wang, N. Zhang, S.P. Lau, W. Zhang, M. Chhowalla, Y. Chai, Topological phase change transistors based on tellurium Weyl semiconductor, Sci. Adv. 8 (2022) eabn3837.
[18] J. Chen, Y. Zhou, J. Yan, J. Liu, L. Xu, J. Wang, T. Wan, Y. He, W. Zhang, Y. Chai, Room-temperature valley transistors for low-power neuromorphic computing, Nat. Commun. 13 (2022) 7758.
[19] C. Rylands, A. Parhizkar, A.A. Burkov, V. Galitski, Chiral anomaly in interacting condensed matter systems, Phys. Rev. Lett. 126 (2021) 185303.
[20] G.W. Semenoff, Condensed-matter simulation of a three-dimensional anomaly, Phys. Rev. Lett. 53 (1984) 2449.
[21] O.E. Parfenov, A.N. Taldenkov, D.V. Averyanov, I.S. Sokolov, O.A. Kondratev, M.M. Borisov, S.N. Yakunin, I.A. Karateev, A.M. Tokmachev, V.G. Storchak, Layer-controlled evolution of electron state in the silicene intercalation compound $SrSi_2$, Mater. Horiz. 9 (2022) 2854.
[22] H. Jia, M. Wang, S. Ma, R.-Y. Zhang, J. Hu, D. Wang, C.T. Chan, Experimental realization of chiral Landau levels in two-dimensional Dirac cone systems with inhomogeneous effective mass, Light: Sci. Appl. 12 (2023) 165.
[23] V.A. Saroka, F. Kong, L. Bogani, C.A. Downing, R.B. Payod, F.R. Fischer, X. Sun, Flat band, tunable chiral anomaly, and pitchfork bifurcation in a honeycomb lattice, Phys. Rev. B 110 (2024) 195134.
[24] M. Wang, Q. Ma, S. Liu, R.-Y. Zhang, L. Zhang, M. Ke, Z. Liu, C.T. Chan, Observation of boundary induced chiral anomaly bulk states and their transport properties, Nat. Commun. 13 (2022) 5916.
[25] H.C. Kao, M. Lewkowicz, B. Rosenstein, Ballistic transport, chiral anomaly, and emergence of the neutral electron-hole plasma in graphene, Phys. Rev. B 82 (2010) 035406.
[26] B. Rosenstein, M. Lewkowicz, T. Maniv, Chiral anomaly and strength of the electron-electron interaction in graphene, Phys. Rev. Lett. 110 (2013) 066602.
[27] S.-Q. Shen, C.-A. Li, Q. Niu, Chiral anomaly and anomalous finite-size conductivity in graphene, 2D Mater. 4 (2017) 035014.
[28] C.-A. Li, Pseudo chiral anomaly in zigzag graphene ribbons, J. Phys.: Condens. Matter 32 (2020) 025301.
[29] V. Saroka, V. Demin, M. Pizzochero, Topological engineering of chiral anomalies in Janus nanoribbons, Nano Lett. 26 (2026) 6532.
[30] S.N. Solodukhin, Manifestations of $(2 + 1)d$ chiral anomaly in a graphene plate, Phys. Lett. B 865 (2025) 139452.
[31] M. Mandal, E. Rha, A. Chotrattanapituk, D.C. Carrizales, A. Lygo, K.B. Woller, M. Cheng, R. Okabe, G. Zhu, S. Kishida, K. Mak, C.-L. Fu, S. Michalopoulos, C. Liu, L. Wu, Y. Zhu, S. Stemmer, M. Li, Tuning chiral anomaly signature in a Dirac semimetal via fast-ion implantation, Nano Lett. 25 (2025) 17571.
[32] T. Iwasaki, S. Nakamura, O.G. Agbonlahor, M. Muruganathan, M. Akabori, Y. Morita, S. Moriyama, S. Ogawa, Y. Wakayama, H. Mizuta, S. Nakaharai, Room-temperature negative magnetoresistance of helium-ion-irradiated defective graphene in the strong Anderson localization regime, Carbon 175 (2021) 87.
[33] Y.-B. Zhou, B.-H. Han, Z.-M. Liao, H.-C. Wu, D.-P. Yu, From positive to negative magnetoresistance in graphene with increasing disorder, Appl. Phys. Lett. 98 (2011) 222502.

[34] B.R. Matis, F.A. Bulat, A.L. Friedman, B.H. Houston, J.W. Baldwin, Giant negative magnetoresistance and a transition from strong to weak localization in hydrogenated graphene, Phys. Rev. B 85 (2012) 195437.
[35] X. Hong, S.-H. Cheng, C. Herding, J. Zhu, Colossal negative magnetoresistance in dilute fluorinated graphene, Phys. Rev. B 83 (2011) 085410.
[36] Y. Zhang, J. You, W. Li, Z. Huang, Y. Feng, Y. Liu, J. Li, Rhenium-induced negative magnetoresistance in monolayer graphene, Magnetochemistry 11 (2025) 39.
[37] N. Briggs, Z.M. Gebeyehu, A. Vera, T. Zhao, K. Wang, A. de la Fuente Duran, B. Bersch, T. Bowen, K.L. Knappenberger, Jr., J.A. Robinson, Epitaxial graphene/silicon carbide intercalation: a minireview on graphene modulation and unique 2D materials, Nanoscale 11 (2019) 15440.
[38] I.S. Sokolov, D.V. Averyanov, O.E. Parfenov, I.A. Karateev, A.N. Taldenkov, A.M. Tokmachev, V.G. Storchak, 2D ferromagnetism in europium/graphene bilayers, Mater. Horiz. 7 (2020) 1372.
[39] S. Schumacher, F. Huttmann, M. Petrović, C. Witt, D.F. Förster, C. Vo-Van, J. Coraux, A.J. Martínez-Galera, V. Sessi, I. Vergara, R. Rückamp, M. Grüninger, N. Schleheck, F. Meyer zu Heringdorf, P. Ohresser, M. Kralj, T.O. Wehling, T. Michely, Europium underneath graphene on Ir(111): Intercalation mechanism, magnetism, and band structure, Phys. Rev. B 90 (2014) 235437.
[40] M. Jugovac, I. Cojocariu, J. Sánchez-Barriga, P. Gargiani, M. Valvidares, V. Feyer, S. Blügel, G. Bihlmayer, P. Perna, Inducing single spin-polarized flat bands in monolayer graphene, Adv. Mater. 35 (2023) 2301441.
[41] I.S. Sokolov, D.V. Averyanov, O.E. Parfenov, A.N. Taldenkov, M.G. Rybin, A.M. Tokmachev, V.G. Storchak, Proximity coupling of graphene to a submonolayer 2D magnet, Small 19 (2023) 2301295.
[42] N.A. Anderson, M. Hupalo, D. Keavney, M. Tringides, D. Vaknin, Intercalated rare-earth metals under graphene on SiC, J. Magn. Magn. Mater. 474 (2019) 666.
[43] N.A. Anderson, M. Hupalo, D. Keavney, M.C. Tringides, D. Vaknin, Intercalated europium metal in epitaxial graphene on SiC, Phys. Rev. Mater. 1 (2017) 054005.
[44] I.S. Sokolov, D.V. Averyanov, O.E. Parfenov, A.N. Mihalyuk, A.N. Taldenkov, O.A. Kondratev, I.A. Eliseyev, S.P. Lebedev, A.A. Lebedev, A.M. Tokmachev, V.G. Storchak, Epitaxial graphene integrated with a monolayer magnet, Carbon 245 (2025) 120794.
[45] X. Qiu, T. Zhu, Z. Fan, K. Wang, Y. Mu, B. Yang, D. Wu, H. Zhang, C. Wang, H. Wang, Y. Zhang, Giant splitting of folded Dirac bands in Kekulé-ordered graphene with Eu intercalation, J. Phys. Chem. Lett. 17 (2026) 2785.
[46] S. Link, S. Forti, A. Stöhr, K. Küster, M. Rösner, D. Hirschmeier, C. Chen, J. Avila, M.C. Asensio, A.A. Zakharov, T.O. Wehling, A.I. Lichtenstein, M.I. Katsnelson, U. Starke, Introducing strong correlation effects into graphene by gadolinium intercalation, Phys. Rev. B 100 (2019) 121407(R).
[47] I.S. Sokolov, D.V. Averyanov, O.E. Parfenov, A.N. Taldenkov, A.M. Tokmachev, V.G. Storchak, Magnetic heterostructure of graphene with a submonolayer magnet on silicon, Carbon 218 (2024) 118769.
[48] I.I. Klimovskikh, M. Krivenkov, A. Varykhalov, D. Estyunin, A.M. Shikin, Reconstructed Fermi surface in graphene on Ir(111) by Gd-Ir surface alloying, Carbon 147 (2019) 182.
[49] M. Kolmer, B. Schrunk, M. Hupalo, J. Hall, S. Chen, J. Zhang, C.-Z. Wang, A. Kaminski, M.C. Tringides, Highly asymmetric graphene layer doping and band structure manipulation in rare earth-graphene heterostructure by targeted bonding of the intercalated gadolinium, J. Phys. Chem. C 126 (2022) 6863.

[50] I.S. Sokolov, D.V. Averyanov, O.E. Parfenov, A.N. Taldenkov, O.A. Kondratev, A.M. Tokmachev, V.G. Storchak, Proximity coupling of graphene to a 2D silicene magnet, Carbon 238 (2025) 120242.
[51] G.J. McCarthy, J.E. Greedan, Complex oxides containing divalent europium. I. Guidelines for the prediction of new phases. Application to phases of the europium metal oxide ($EuMO_3$), Inorg. Chem. 14 (1975) 772.
[52] A.M. Tokmachev, D.V. Averyanov, A.N. Taldenkov, I.S. Sokolov, I.A. Karateev, O.E. Parfenov, V.G. Storchak, Two-dimensional magnets beyond the monolayer limit, ACS Nano 15 (2021) 12034-12041.
[53] I. El Hajj, L. Speyer, S. Cahen, P. Lagrange, G. Medjahdi, C. Hérold, Crystal structure of first stage strontium-graphite intercalation compound, Carbon 168 (2020) 732.
[54] H. Silotia, A. Kumari, A. Gupta, J. De, S.K. Pal, R. Tomar, S. Chakraverty, Possible signatures of chiral anomaly in the magnetoresistance of a quasi-2-dimensional electron gas at the interface of $LaVO_3$ and $KTaO_3$, Adv. Electron. Mater. 8 (2022) 2200195.
[55] A.A. Lebedev, S.Y. Davydov, I.A. Eliseyev, A.D. Roenkov, O. Avdeev, S.P. Lebedev, Y. Makarov, M. Puzyk, S. Klotchenko, A.S. Usikov, Graphene on SiC substrate as biosensor: Theoretical background, preparation, and characterization, Materials 14 (2021) 590.
[56] J. Zhao, P. Ji, Y. Li, R. Li, K. Zhang, H. Tian, K. Yu, B. Bian, L. Hao, X. Xiao, W. Griffin, N. Dudeck, R. Moro, L. Ma, W.A. de Heer, Ultrahigh-mobility semiconducting epitaxial graphene on silicon carbide, Nature 625 (2024) 60.
[57] D. Yang, F. Ma, X. Bian, Q. Xia, K. Xu, T. Hu, The growth of epitaxial graphene on SiC and its metal intercalation: a review, J. Phys.: Condens. Matter 36 (2024) 173003.
[58] Y.-M. Lin, A. Valdes-Garcia, S.-J. Han, D.B. Farmer, I. Meric, Y. Sun, Y. Wu, C. Dimitrakopoulos, A. Grill, P. Avouris, K.A. Jenkins, Wafer-scale graphene integrated circuit, Science 332 (2011) 1294.
[59] N. Briggs, B. Bersch, Y. Wang, J. Jiang, R.J. Koch, N. Nayir, K. Wang, M. Kolmer, W. Ko, A. de la Fuente Duran, S. Subramanian, C. Dong, J. Shallenberger, M. Fu, Q. Zou, Y.-W. Chuang, Z. Gai, A.-P. Li, A. Bostwick, C. Jozwiak, C.-Z. Chang, E. Rotenberg, J. Zhu, A.C.T. van Duin, V. Crespi, J.A. Robinson, Atomically thin half-van der Waals metals enabled by confinement heteroepitaxy, Nat. Mater. 19 (2020) 637.
[60] W.A. de Heer, C. Berger, X. Wu, M. Sprinkle, Y. Hu, M. Ruan, J.A. Stroscio, P.N. First, R. Haddon, B. Piot, C. Faugeras, M. Potemski, J.-S. Moon, Epitaxial graphene electronic structure and transport, J. Phys. D: Appl. Phys. 43 (2010) 374007.
[61] D.T. Son, B.Z. Spivak, Chiral anomaly and classical negative magnetoresistance of Weyl metals, Phys. Rev. B 88 (2013) 104412.
[62] A.A. Burkov, Chiral anomaly and diffusive magnetotransport in Weyl metals, Phys. Rev. Lett. 113 (2014) 247203.
[63] H. Zhang, M. Wu, Y. Han, X. Zhu, J. Zhou, W. Ning, Competition between chiral anomaly and weak antilocalization in $Cd_3As_2$ nanoplates, J. Solid State Chem. 333 (2024) 124628.
[64] B. Zhao, P. Cheng, H. Pan, S. Zhang, B. Wang, G. Wang, F. Xiu, F. Song, Weak antilocalization in $Cd_3As_2$ thin films, Sci. Rep. 6 (2016) 22377.
[65] N.P. Armitage, E.J. Mele, A. Vishwanath, Weyl and Dirac semimetals in three-dimensional solids, Rev. Mod. Phys. 90 (2018) 015001.
[66] Z. Jia, C. Li, X. Li, J. Shi, Z. Liao, D. Yu, X. Wu, Thermoelectric signature of the chiral anomaly in $Cd_3As_2$, Nat. Commun. 7 (2016) 13013.
[67] C.-L. Zhang, S.-Y. Xu, I. Belopolski, Z. Yuan, Z. Lin, B. Tong, G. Bian, N. Alidoust, C.-C. Lee, S.-M. Huang, T.-R. Chang, G. Chang, C.-H. Hsu, H.-T. Jeng, M. Neupane, D.S. Sanchez, H. Zheng, J. Wang, H. Lin, C. Zhang, H.-Z. Lu, S.-Q. Shen, T. Neupert, M.Z. Hasan, S. Jia,

Signatures of the Adler-Bell-Jackiw chiral anomaly in a Weyl fermion semimetal, Nat. Commun. 7 (2016) 10735.
[68] M. Behnami, D.V. Efremov, S. Aswartham, G. Shipunov, B.R. Piening, C.G.F. Blum, V. Kocsis, J. Dufouleur, I. Pallecchi, M. Putti, B. Büchner, H. Reichlova, F. Caglieris, Signature of chiral anomaly in the Weyl semimetal $TaRhTe_4$, Phys. Rev. B 112 (2025) 045101.

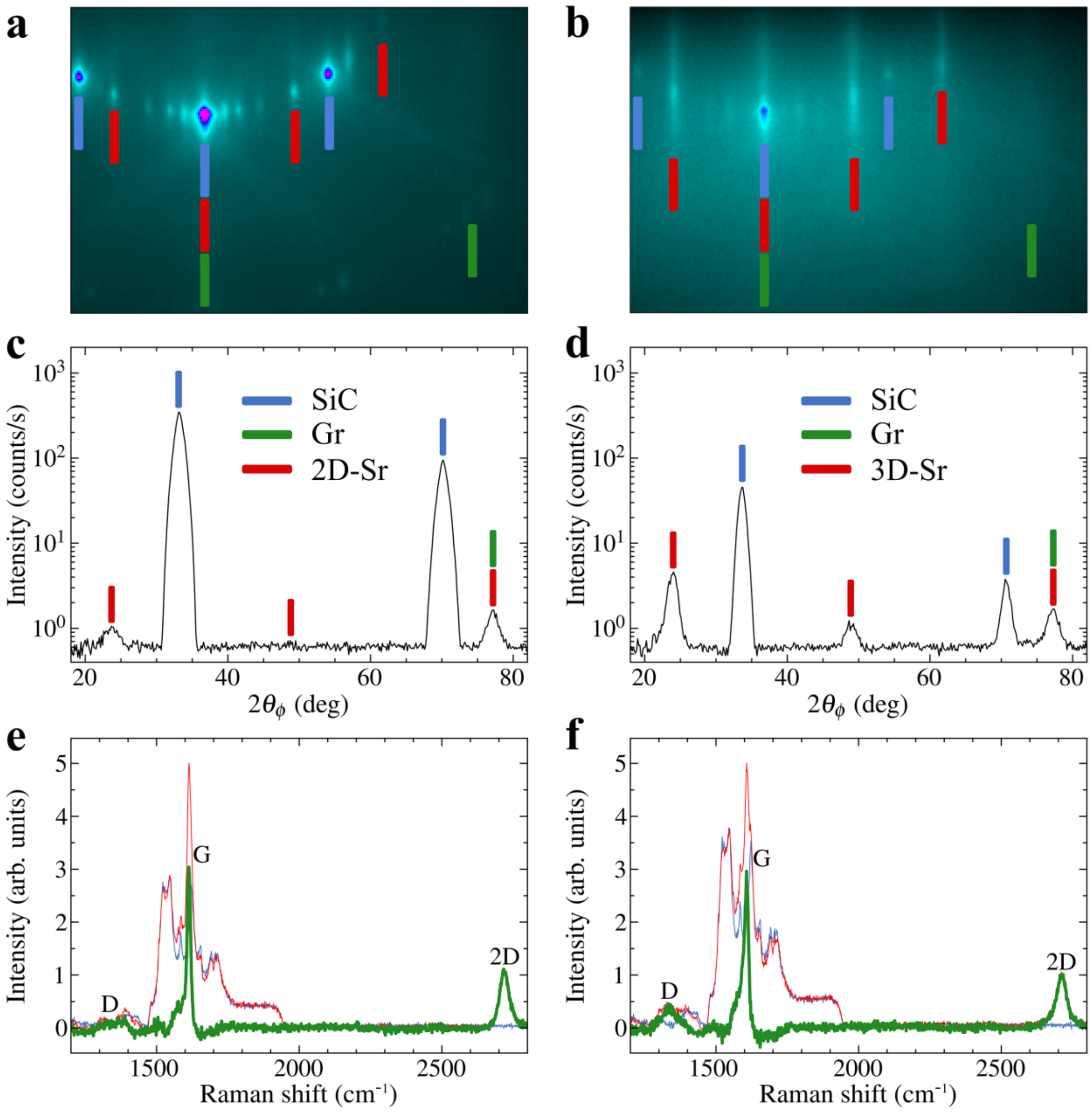

**Fig. 1.** a) and b) RHEED images (along the azimuth [$\bar{1}\bar{1}20$]) of 2D- and 3D-Gr/Sr/SiC, respectively; the reflexes are indicated for SiC (blue), Gr (green), and Sr (red). c) and d) GIXRD scans along the direction [$1\bar{1}00$] of the substrate of 2D- and 3D-Gr/Sr/SiC, respectively; the reflexes are indicated for SiC (blue), Gr (green), and Sr (red). e) and f) Raman spectra of 2D- and 3D-Gr/Sr/SiC, respectively; the raw data spectrum (red), the substrate spectrum (blue) and their difference (green).

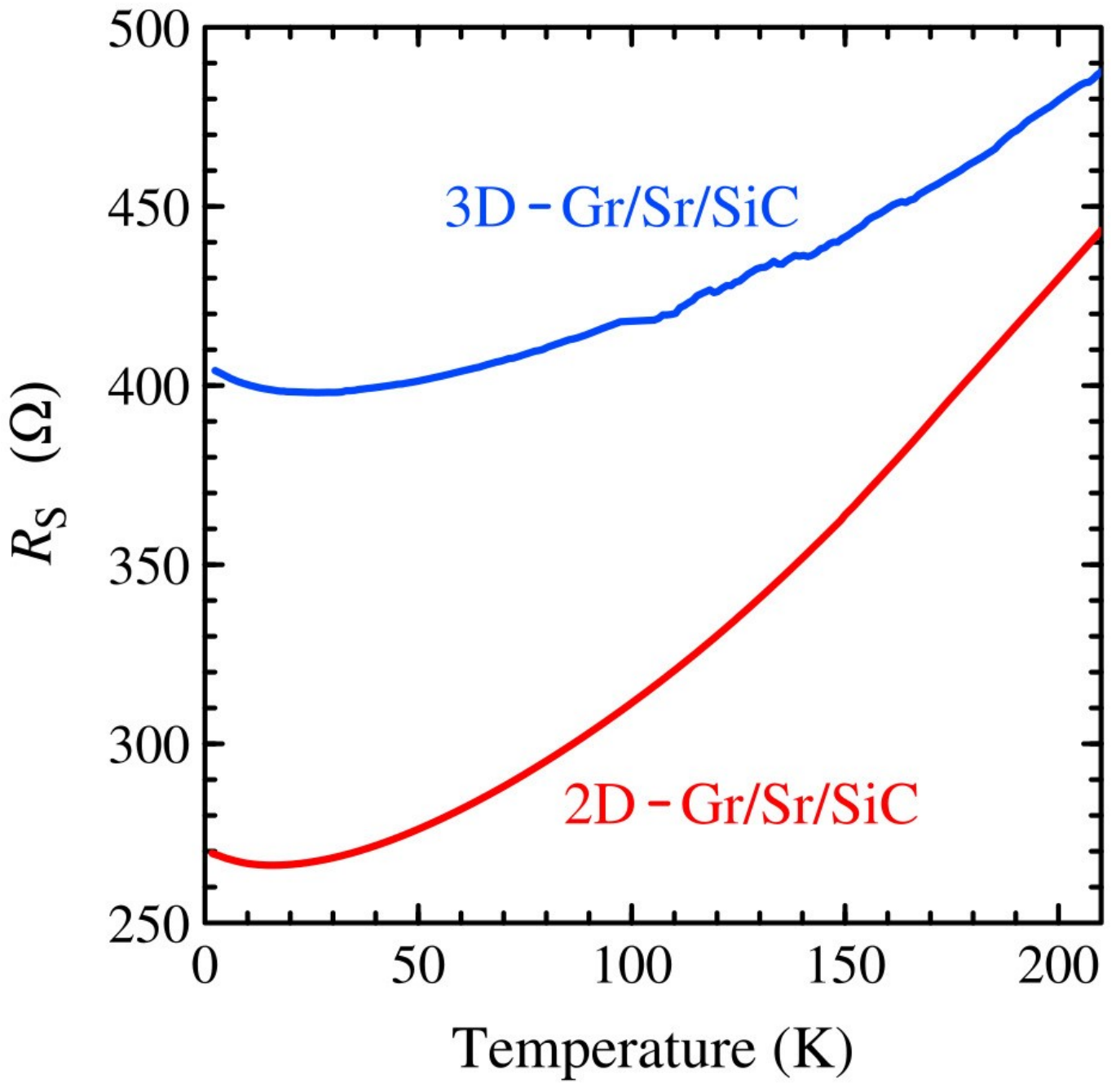

**Fig. 2.** Temperature dependence of sheet resistance in 2D-Gr/Sr/SiC (red) and 3D-Gr/Sr/SiC (blue).

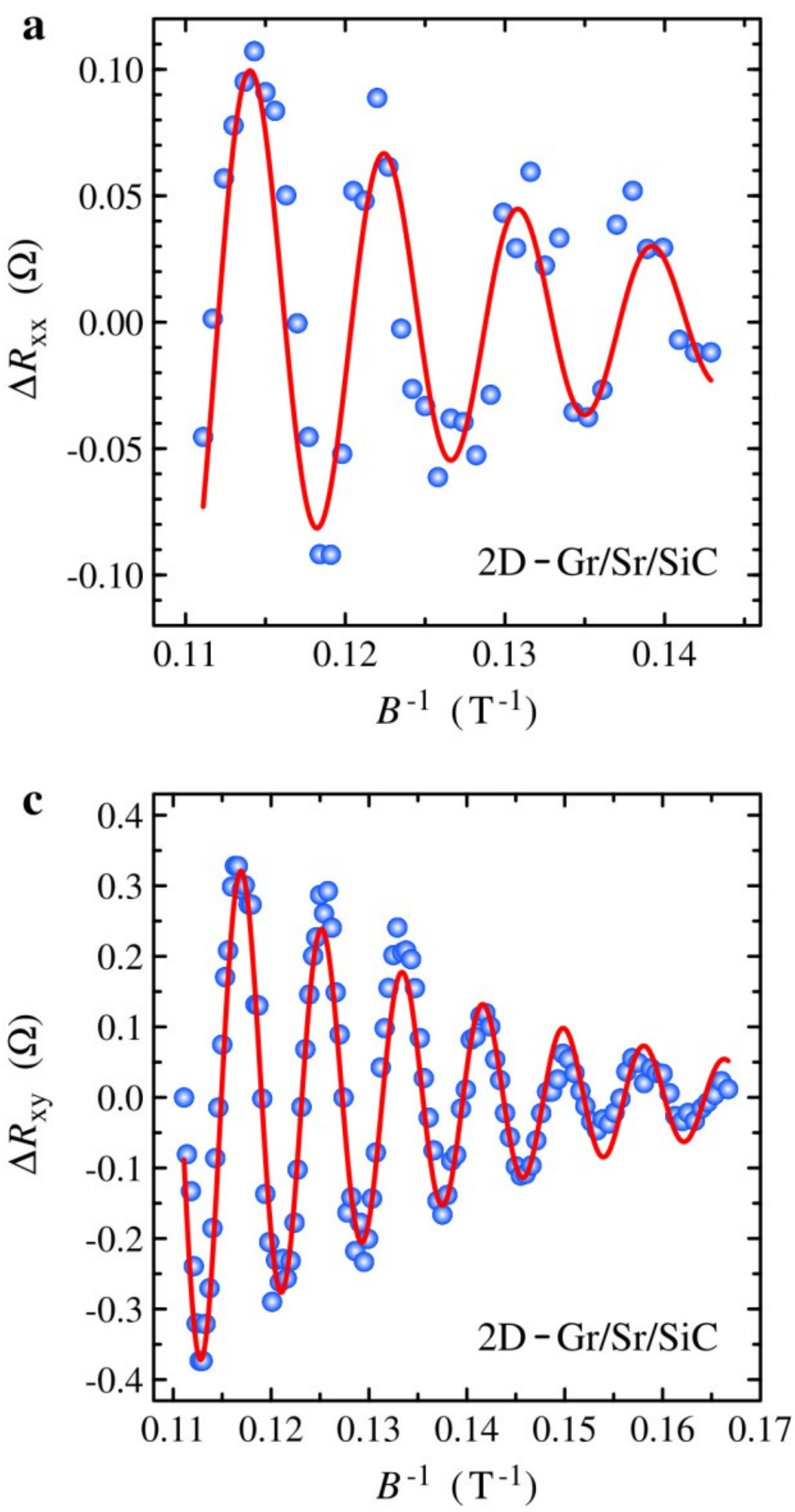

**Fig. 3.** Oscillatory contributions to the sheet resistance (a, b) and the Hall resistance (c, d) in 2D-Gr/Sr/SiC (a, c) and 3D-Gr/Sr/SiC (b, d) at 1.8 K.

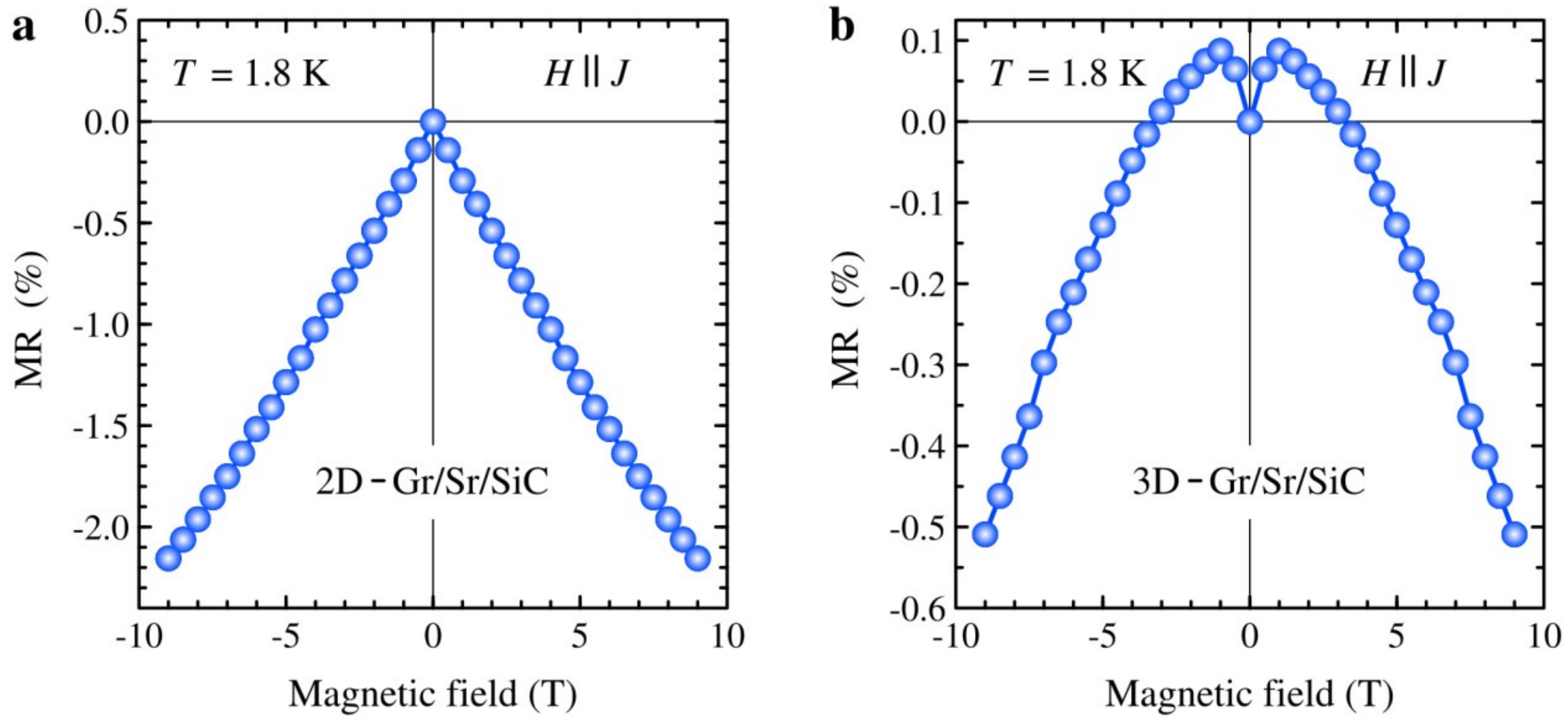

**Fig. 4.** Dependence of magnetoresistance at 1.8 K on magnetic field parallel to the current in a) 2D-Gr/Sr/SiC and b) 3D-Gr/Sr/SiC.

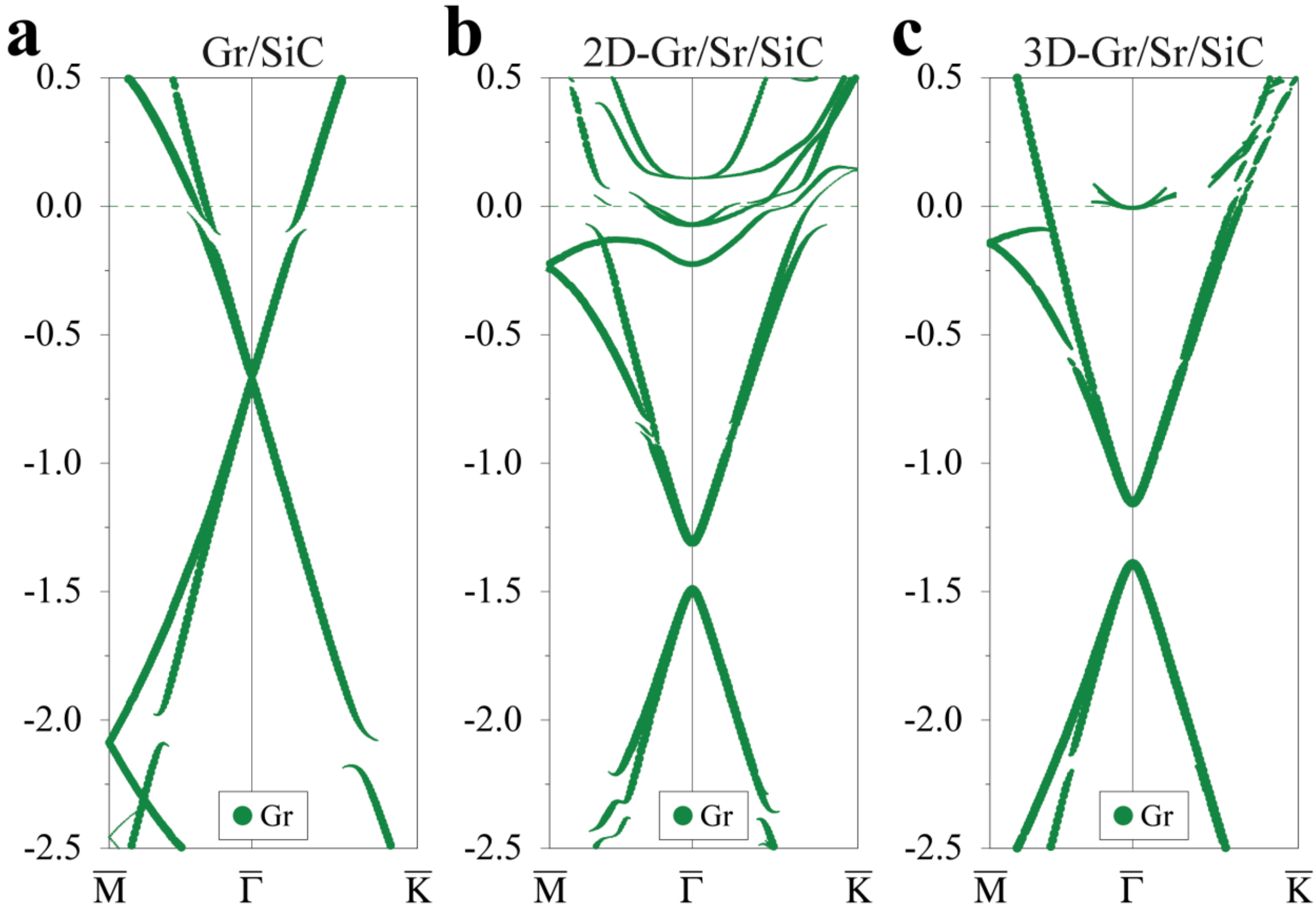

**Fig. 5.** Band structures of a) Gr/SiC, b) 2D- and c) 3D-Gr/Sr/SiC.

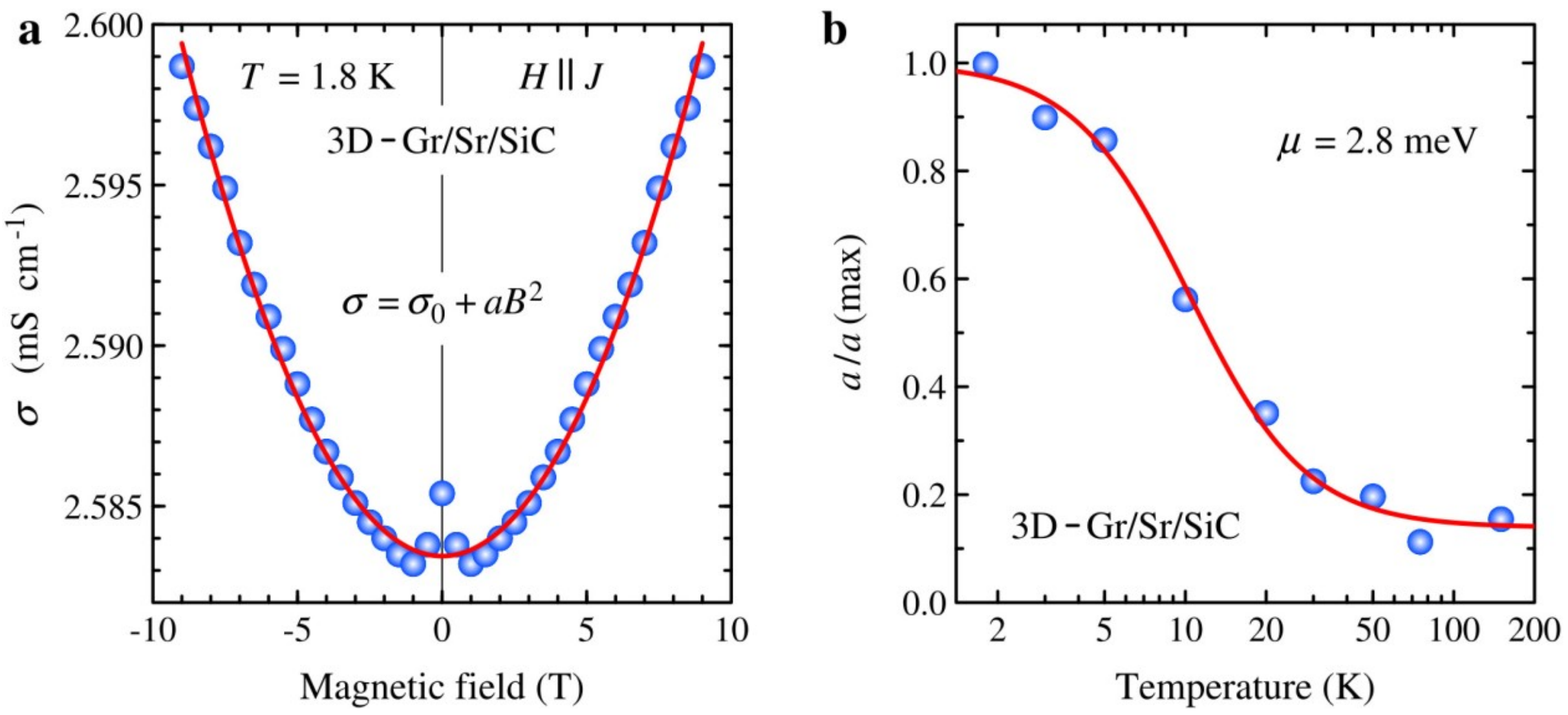

**Fig. 6.** a) Dependence of 3D-Gr/Sr/SiC conductivity at 1.8 K on magnetic field parallel to the current: experimental data (blue dots) and their quadratic approximation $\sigma = \sigma_0 + aB^2$ (red line). b) Temperature dependence of the normalized coefficient $a$: experimental data (blue dots) and approximation of their variation by a function proportional to $(T^2 + \mu^2/\pi^2)^{-1}$ (red line); the chiral chemical potential $\mu$ is estimated at 2.8 meV.